\documentclass[conference,letter]{IEEEtran}
\IEEEoverridecommandlockouts
\usepackage{cite}
\usepackage{amsmath,amssymb,amsfonts}
\usepackage{algorithmic}
\usepackage{graphicx}
\usepackage{textcomp}
\usepackage{xcolor}
\def\BibTeX{{\rm B\kern-.05em{\sc i\kern-.025em b}\kern-.08em
    T\kern-.1667em\lower.7ex\hbox{E}\kern-.125emX}}
\begin{document}

\title{A journey in applying blockchain for cyberphysical systems}

\author{Volkan Dedeoglu,
        Ali Dorri,
        Raja Jurdak,
	Regio A. Michelin,
	Roben C. Lunardi,
    Salil S. Kanhere,
        and~Avelino F. Zorzo%
\thanks{V. Dedeoglu is with CSIRO Data61, Pullenvale, Australia e-mail: name.surname@csiro.au}%
\thanks{A. Dorri and R. Jurdak are with Queensland University of Technology, Australia email: \{r.jurdak, ali.dorri\}@unsw.edu.au}%
\thanks{R.A. Michelin is with Cyber Security CRC and UNSW Sydney, Australia email: name.surname@cybersecuritycrc.org.au}%
\thanks{S.S. Kanhere is with UNSW Sydney, Australia email: name.surname @unsw.edu.au}%
\thanks{R.C. Lunardi and A.F. Zorzo are with Pontificia Universidade Catolica do Rio Grande do Sul, Brazil email: name.surname@pucrs.br}
}

\maketitle

\begin{abstract}
Cyberphysical Systems (CPS) are transforming the way we interact with the physical world around us. However, centralised approaches for CPS systems are not capable of addressing the unique challenges of CPS due to the complexity, constraints, and dynamic nature of the interactions. To realize the true potential of CPS, a decentralized approach that takes into account these unique features is required. Recently, blockchain-based solutions have been proposed to address CPS challenges. Yet, applying blockchain for diverse CPS domains is not straightforward and has its own challenges. In this paper, we share our experiences in applying blockchain technology for CPS to provide insights and highlight the challenges and future opportunities.      
\end{abstract}

\begin{IEEEkeywords}
blockchain, IoT, CPS
\end{IEEEkeywords}

\section{Introduction}
Cyberphysical Systems (CPS) connect the physical and cyber worlds by integrating sensing, networking, computing, decision making, and actuation processes with a potential to transform the way we interact with the physical world around us~\cite{Rajkumar2010}. However, seamless integration is challenging due to the complexity, constraints, and dynamic nature of the interactions in CPS. The  prominent CPS challenges that remain to be addressed include the lack of central control, resource constrained and heterogeneous devices, scale of the network and the data generated, lack of trust between devices, dynamic network structure, privacy and security risks.

Blockchain is a promising technology to address the aforementioned CPS challenges with its salient features of decentralization, anonymity, and security~\cite{BlockchainInIoT}. Introduced in 2008 as the technology underpinning the Bitcoin~\cite{Satoshi2008} cryptocurrency network, blockchain is an immutable and distributed ledger that stores data in blocks chained together with cryptographic mechanisms in a peer-to-peer network. Although the initial Bitcoin blockchain's main functionality is to transfer coins between trustless network participants without requiring any trusted intermediaries, blockchains with different functionalities have since evolved to support other applications~\cite{Angelis}. For example, Ethereum and Hyperledger blockchains support smart contract functionality to run self-executing programs when predefined conditions are met~\cite{Wang2018}.

\begin{figure}[h]
    \centering
    \includegraphics[width=3.1in]{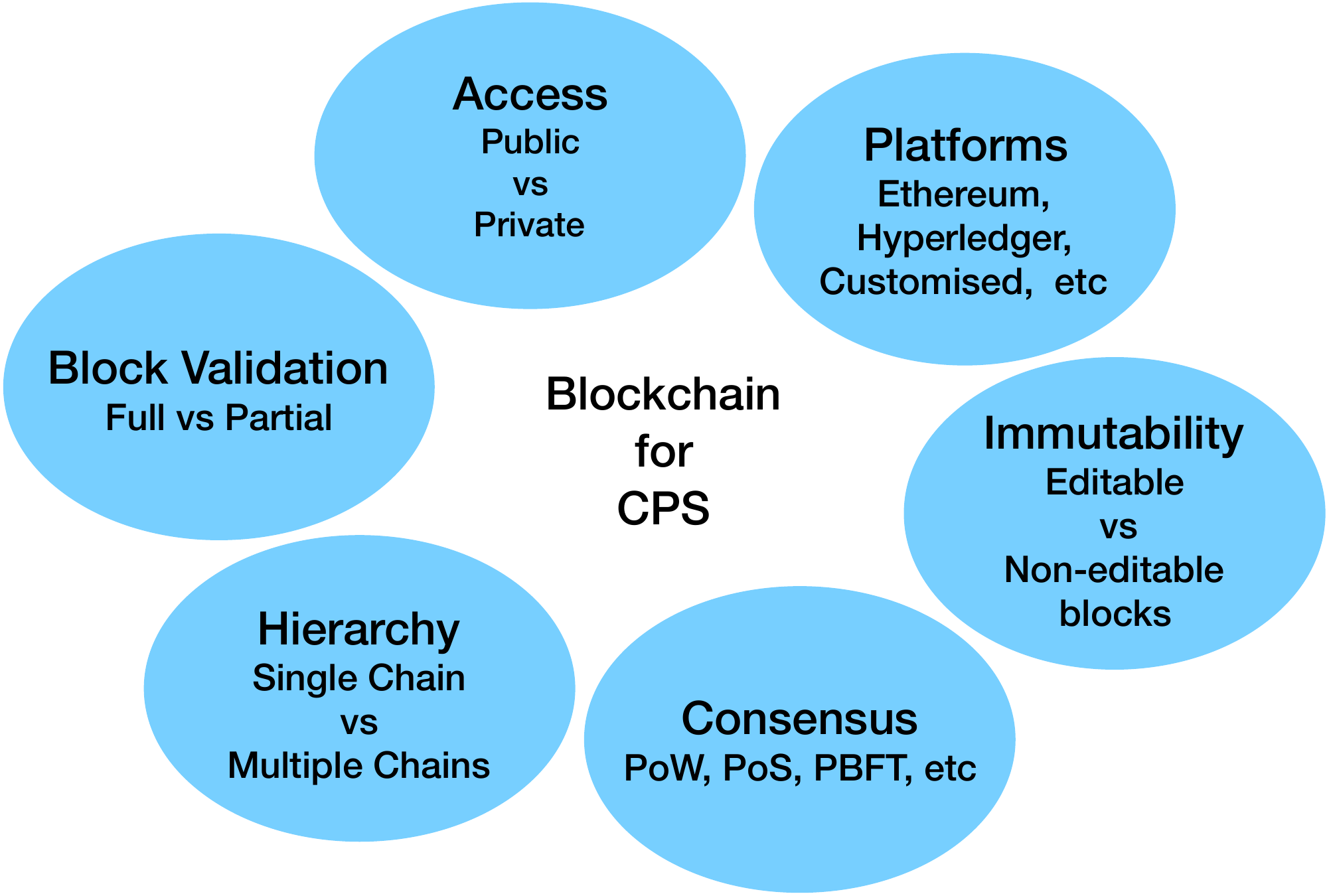}
    \caption{The design choices for blockchain for CPS.}
    \label{fig:designChoices}
    \vspace{-0.5cm}
\end{figure}

Although blockchain has favorable features for CPS applications, adopting blockchain in CPS is not straightforward mainly due to scalability issues, high resource consumption, transaction latency, low throughput, privacy issues, and lack of trust. Understanding how different blockchain architectures work can help us choose the most appropriate blockchain structure for a given application~\cite{DependableIoT}. CPS have diverse application domains due to their unique features~\cite{Khaitan2015}, and as a result of the application-specific constraints and requirements, there is no one-size-fits-all solution for the design of blockchain-based solutions for CPS. Fig.\ref{fig:designChoices} presents some of the design choices for blockchain-based CPS solutions.   

\begin{figure}[t]
    \centering
    \includegraphics[width=3.4in]{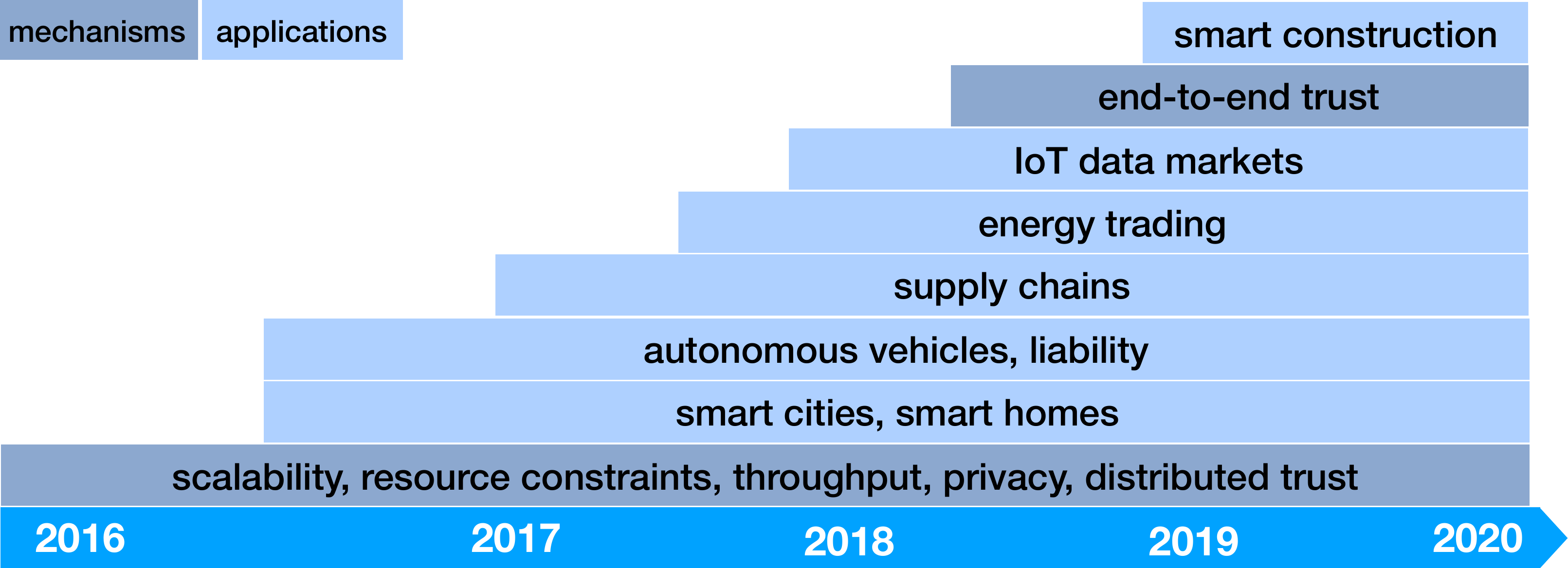}
    \caption{Research road map}
    \label{fig:roadMap}
\end{figure}

In this paper, we share our experiences in applying blockchain technology for CPS to provide insights and highlight the challenges and future opportunities. Fig.~\ref{fig:roadMap} shows the roadmap covering our research in blockchain for CPS starting from year 2016. As presented in the figure, we can broadly categorize our research work in two main categories: (1) developing blockchain mechanisms to address common CPS challenges, and (2) building blockchain-based solutions for different CPS application domains including smart cities, smart homes, autonomous vehicles, supply chains, energy trading, IoT data markets, and smart construction. 

The rest of this paper is organized as follows: Section II presents the common challenges involved in applying  blockchain technology for CPS and the mechanisms that we proposed to address these challenges. Section III presents our experience in building blockchain-based solutions for various CPS applications. In Section IV, we discuss our research approach, provide insights about applying blockchain for CPS, and highlight current problems and limitations for blockchain adoption for CPS.     

\section{Designing blockchain for CPS}
Blockchain has received tremendous attention as a means to provide security, anonymity, auditability, and trust. However, the existing blockchain-based solutions suffer from a number of challenges  and thus are not directly applicable in CPS. In the next sections, we discuss the fundamental challenges involved in applying blockchain for CPS and briefly describe the mechanisms that we proposed to address these challenges as summarized in Table~\ref{tbl:BC_design}.  

\subsection{Scalability and resource constraints}
Conventional blockchains consume significant resources of the participating nodes, in particular the miners, to manage the blockchain. This is mainly due to: i) Consensus algorithm, which requires the miners to solve puzzles or provide proof of X before mining a new block, ii) broadcasting transactions and blocks, which in turn increases the packet overhead and bandwidth consumption in the participating nodes, iii) verifying new blocks, which increases the processing time at all particapting nodes for verifying all transactions in newly arrived blocks,  iv) blockchain immutability, which increases the size of the blockchain database as all transactions and blocks must be stored permanently by the participating nodes. The large scale of CPS amplifies the outlined challenges and limits the blockchain applicability. \par 
In \cite{LSB} we proposed a hierarchical blockchain-based framework to address the aforementioned challenges. We proposed a Distributed Time-based Consensus (DTC) that requires the miners to wait for a particular time period before storing a block, which significantly reduces the mining overhead. To reduce the packet overhead associated with broadcasting transactions in the blockchain, we proposed to cluster the underlying peer-to-peer network and only the Cluster Heads (CHs) manage the blockchain by verifying and storing new transactions and blocks. We proposed a distributed trust algorithm where the CHs (i.e., the miners) gradually build up trust on each other. Based on the trust level, the CHs verify less portion of transactions in the newly mined blocks by each CH, which in turn reduces the processing overhead for verifying new blocks. 

Another scalability challenge with conventional blockchains is throughput, which  is defined as the total number of transactions that can be stored in the blockchain per second.  In CPS, large number of devices and users that frequently interact with one another, generate significant number of transactions. This potentially demands high throughput for blockchain. However, conventional blockchain instantiations suffer from limited throughput, e.g., Bitcoin throughput is 7 transactions per second and Ethereum throughput is approximately 15 transactions per second. The throughput is  managed by the consensus algorithm, e.g., POW adjusts the difficulty in solving the cryptographic puzzle  in a way that only one block can be generated in each 10 minutes, which in turn limits the blockchain throughput. 
Our LSB~\cite{LSB} approach addresses scalability through a Distributed Throughput Management (DTM) algorithm. Recall that  \cite{LSB}  introduces DTC and only CHs manage the blockchain. DTM tunes two parameters, which are the time period in which CHs can store a new block and the number of CHs in the network to ensure that the network throughput does not deviate significantly from the network load.  In \cite{MOF-BC} we proposed a solution that enables   users to remove transactions from the blockchain while maintaining  consistency and thus reduces the size of the blockchain. Unlike conventional blockchain instantiations where the block hash is calculated by hashing all block content, in   \cite{MOF-BC} the block hash is calculated by hashing the transaction IDs, thus enabling the users to remove transaction content while maintaining the transaction ID that ensures consistency. 

\subsection{Privacy}
In a blockchain all participating nodes are known by a Public Key (PK), which can be changed for generating new transactions, which in turn increases user anonymity. However, studies in Bitcoin blockchain show that the attackers can attempt to deanonymize a user by linking multiple transactions in the blockchain with a real-world identity, which in turn compromises the user privacy \cite{LSB}. \par 
In CPS setting, all interactions between the participating nodes are permanently recorded in  blockchain, which can be accessed by the participating nodes. This potentially introduces high privacy concerns for the CPS users as the attackers can have access to the full history of the interactions of the users since joining the blockchain \cite{Privacy-Preserving}. Similar to Bitcoin, to protect user privacy in CPS settings the user can change his PK. However, this method is vulnerable to linking attack. The users may  employ multiple ledgers for storing their transactions that splits the history of the transactions of the user \cite{Privacy-Preserving}. This potentially reduces the success rate of linking attack. However, generating new ledgers increases the costs incurred to the end user. Thus, the user needs to consider the trade-off between cost and privacy. \par
As outlined above, the immutability feature of blockchain makes it impossible to remove any transaction from the blockchain. However, one of the key requirements of privacy regulations, including General Data Protection Regulation (GDPR) of the EU \cite{GDPR_reg}, is for the users to have the right to experience the right for their data to be forgotten. Blockchain users may attempt to store illegal content in the blockchain \cite{Illegal2019}, which is permanently stored in the blokchain and publicly available to all users. With the existing blockchain-based solutions, removing a transaction (or part of a transaction) changes the hash of the corresponding block, which introduces inconsistency in the blockchain. \par 
In  \cite{MOF-BC} we proposed a memory optimized blockchain framework that enables IoT users to remove or summarize transactions or store temporarily transactions in the blockchain and thus reduces the history of transactions in the blockchain and enhances user privacy. The proposed method also enables the users to experience the right for their data to be forgotten and thus increases blockchain compatibility with GDPR. In \cite{Privacy-Preserving} we studied the success rate in classifying IoT devices based on the history of transactions in the blockchain using Machine Learning (ML) algorithms. Note that there is no blockchain instantiation that is currently being used braodly by IoT devices. Thus, we populated a blockchain based on the traffic of IoT devices in a smart home available in \cite{SmartHomeDataset}. The dataset contains traffic data of 28 IoT devices in a smart home setting. The results show that the attacker can identify the devices with 90\% success rate. We proposed multiple obfuscation methods to reduce the success rate, which includes delaying transactions, combining transaction in a ledger, and using multiple ledgers to chain transactions of a device.

\subsection{Trust}
Due to the intertwined nature of the physical and computational elements in CPS, improving the trust in the system is a highly challenging problem. Typical CPS may involve a network of heterogeneous and trustless entities interacting with each other and participating in sensing, networking, storage and manipulation of data, and actuation tasks. While improving the trust in the system, these tasks and the interactions should be taken into account. 

CPS rely on the sensing data collected from the physical world. Storing this data on a blockchain ensures the immutability of data. Thus, entities trust that the data cannot be altered or removed without getting detected after it is recorded on a blockchain. However, blockchain mechanisms cannot guarantee the trustworthiness of data at the origin, as the data is collected from the physical world by an entity that may be malicious or erroneous. To address the data trust challenge, we proposed a trust architecture that uses multiple sensor observations to evaluate the trustworthiness of sensing data in~\cite{TrustArchitecture}. 

When there are multiple data sources, they can be used to establish trust in data provided by the entities and ensure that the data recorded on the blockchain is trusted. In~\cite{TrustChain}, we proposed a trust and reputation mechanism to address the issue of trust associated with the quality of commodities and the entities providing data to the blockchain in supply chains.

Furthermore, the participating nodes in the blockchain are anonymous nodes that do not trust each other. Thus, all transactions in the newly mined blocks must be verified by all participating nodes, which in turn increases the processing overhead in the miners. In Lightweight Scalable Blockchain (LSB)~\cite{LSB}, we proposed a distributed trust mechanism that decreases the processing overhead for validating new blocks based on the trust blockchain nodes build towards each other through direct or indirect evidence. Our end-to-end trust architecture in~\cite{TrustArchitecture} also adapts block verification based on the reputation of the block generating nodes.

\begin{table}
  \caption{CPS challenges and proposed blockchain mechanisms}
  \label{tbl:BC_design}
  \includegraphics[width=\linewidth]{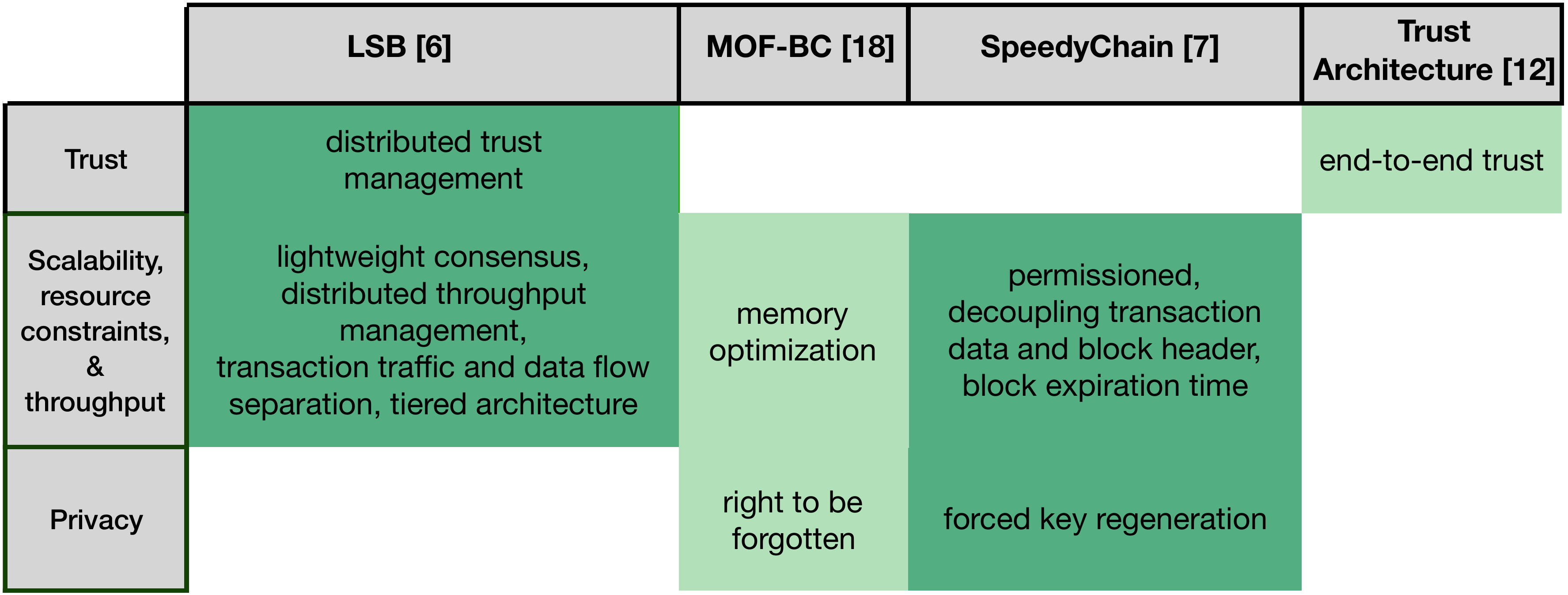}
\end{table}

\section{CPS Applications}
In the previous section, we state the common challenges related to applying blockchain technology for CPS and our proposed blockchain mechanisms to address these challenges. In this section, focusing on different application domains, we identify the unique challenges, and present our solutions and the lessons learned as summarized in Table~\ref{tbl:CPSapplications}.    
\begin{table*}
  \caption{CPS applications and proposed blockchain solutions}
  \label{tbl:CPSapplications}
  \centering
  \includegraphics[width=5.7in]{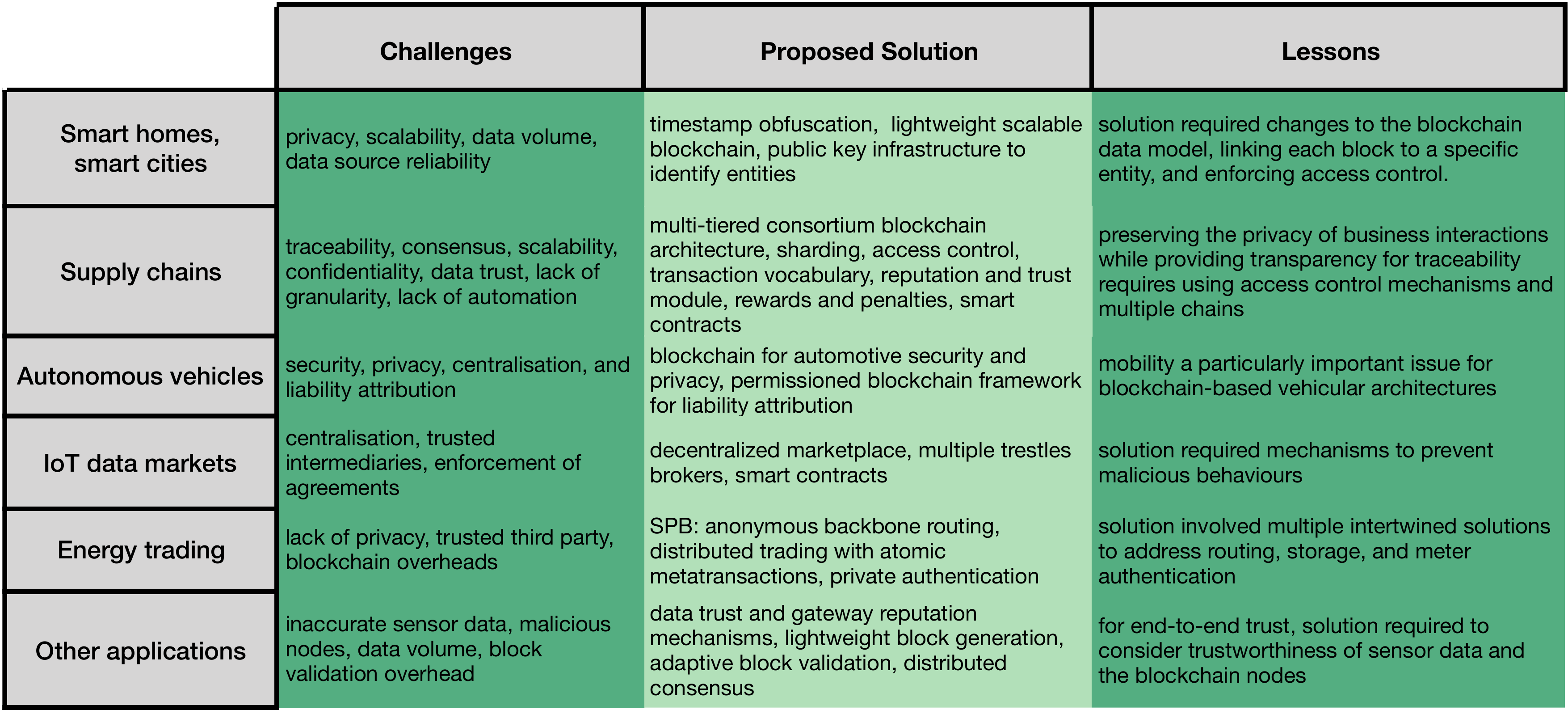}
\end{table*}

\subsection{Smart homes, smart cities}
The number of connected devices present in each home is increasing. These devices are introduced in the home context to improve day-by-day activities making the life of the residents more comfortable. Additionally, these devices are capable of taking decisions based on collected and received information, which makes them smart. In a larger scale, the data collated from a smart home becomes an important piece of information within the context of a smart community and city. A smart city is not limited to smart homes, but also involves entities such as vehicles, service providers, public authorities. The management of data and entities for this scenario presents new opportunities and challenges, as discussed next. 

\paragraph{Challenges}
In a smart home, a resident can use multiple devices, each presenting a different set of characteristics, many of which are resource constrained and from different vendors. Ensuring access control for heterogeneous devices poses a security challenge. This type of challenge must be taken into account when defining IoT solutions in different domains, from smart homes to smart cities. Among the challenges that a smart city presents, we highlight the scalability and the data volume produced from  different sources, for example, Intel predicts that future vehicles will produce 4,000GB of data every day~\cite{Krzanich}. Additionally, the information must be reliable delivered with low latency, as it acts as input to critical systems that make decisions such as intelligent transportation systems, and reporting traffic incidents. Thus it is vital to ensure that the data was not tampered and produced from a trusted source.

\paragraph{Solutions}
Using the LSB~\cite{LSB} we define a decentralised solution forming an overlay network where high resource devices jointly manage a public blockchain that ensures end-to-end privacy and security. LSB relies on the Overlay Block Managers (OBM), which are responsible for maintaining the blockchain and a key list to enforce the device access control.

To address the reliability and latency of the smart city data, we proposed the SpeedyChain~\cite{SpeedyChain} framework. SpeedyChain relies on the public key infrastructure (PKI) for identifying each smart city entity by its public key. Each entity has one block active per public key, in which the produced information is stored. Each block can only store transactions produced from the entity, which owns the private key associated to that block. This blockchain data model enables SpeedyChain to parallelize the information appended to the blockchain, as different entities can add data transactions in their respective block at the same time, leading the solution to provide low-latency data management. 

\paragraph{Lessons Learned}
As a result of our research, we identify that the blockchain architecture solution is capable of handling a large data volume produced in a smart city and still ensures the trust in the information producers. Achieving this goal, however, required changes to the blockchain data model, such as decoupling transactions, linking each block to a specific entity, and using OBMs to enforce access control.

\subsection{Supply chains}
The interconnected and complex structure of today's supply chains transcends geographical boundaries and is shaped by consumer demands, continuous innovation for improved efficiency and integration, and globalisation. While legacy supply chain management systems were effective solutions for simple supply chains, their ability to address the challenges of modern supply chains is limited. Blockchain is an emerging technology that has the potential to revolutionise  supply chains by addressing the key challenges described below.  

\paragraph{Challenges}
As supply chains get more complicated and geographically dispersed, it becomes harder to track the flow of assets, products, information, and payments among supply chain agents. Combined with the increased consumer demand towards the provenance of products, traceability and transparency have become major challenges of modern supply chains. Current supply chains also suffer from scattering of supply chain data across multiple isolated data silos. To build a product story for provenance, the information should be collated from these isolated data silos. Furthermore, supply chain data is susceptible to inaccurate and erroneous data recording, and data tampering. Thus, there is a need for designing mechanisms that improve the integrity of supply chain data and trustworthiness of supply chain agents. Finally, manually executed supply chain processes result in lower system performance in terms of efficiency and costs.

\paragraph{Solutions}
Using food supply chains as an example to address the traceability problem, we proposed ProductChain~\cite{ProductChain}, which is a scalable blockchain framework governed by a consortium of food supply chain entities (e.g., farmers, producers, transporters, retailers, government, and regulatory bodies, etc.). The multi-tiered architecture of ProductChain allows consumers and stakeholders to access the product provenance information while preserving the confidentiality of trade flows by controlling the data access rights. To improve the scalability, the architecture is built on a set of parallel blockchains (shards) that store different types of information and interactions based on our transaction vocabulary.

ProductChain solves the supply chain traceability problem by creating an immutable record of product data and supply chain interactions. However, the data trust problem related to the quality of products and the supply chain participants recording data on the blockchain was not considered. Next, we addressed this supply chain trust problem and proposed the TrustChain~\cite{TrustChain} architecture as a blockchain-based trust management framework. Built on data, blockchain, and application layers, our framework evaluates the truthfulness of supply chain data based on multiple data sources. The reputation and trust module of TrustChain provides granular reputation scores for products and supply chain participants using smart contracts that automate the process. The proposed framework uses these reputation scores to reward or penalize the supply chain participants.

\paragraph{Lessons Learned}
The adoption of blockchain technology by supply chains depends on the performance and overhead of the proposed solutions. The implementations of ProductChain and TrustChain on Hyperledger Fabric\footnote{https://www.hyperledger.org/projects/fabric} demonstrated the effectiveness of the proposed blockchain-based traceability and trust management architectures. Another important challenge for blockchain-based supply chain applications is the tradeoff between transparency and privacy. Our architectures preserve the privacy of business transactions and interactions while providing transparency for product traceability using access control mechanisms and multiple chains.

\subsection{Connected and autonomous vehicles}
As vehicles become increasingly connected and incorporate autonomy features, new opportunities for traffic control, road safety and novel vehicular services such as car sharing or electric vehicle charging. This rapid development in connectedness and autonomy also brings new challenges, which we discuss next. 
\paragraph{Challenges}
Vehicle connectivity is coupled with new security and privacy concerns that range from scalability due to centralisation of current vehicular communication models, safety critical risks if the vehicle control units are fully compromised by an attacker, to exposing the owner's data and activities.  Centralisation is a bottleneck due to the large number of vehicles and the central nodes becoming single points of failure. Safety critical risks range from modifying the states of vehicles control units remotely to fully controlling the vehicle by remote attackers. Privacy risks arise as the vehicle owner accesses services through the vehicle's communication interface, such as wireless remote service updates, when attackers successfully access or modify the messages to infer the owner's identity, activities, or whereabouts. Increased connectivity and autonomy also give rise to new challenges regarding liability assignment when vehicles have an accident, as there are multiple potentially liable entities, such as manufacturer, software provider, service technician, or vehicle owner that may be able to access and alter the state of the vehicles' data to evade liability. 
\paragraph{Solutions}
We address the security and privacy concerns of connected vehicles by tailoring our LSB architecture of the vehicular ecosystem \cite{BCAutomotive2019}, where an overlay of available and computationally capable nodes manages the blockchain using the distributed time-based consensus algorithm. Vehicles can dynamically associate with their closest overlay node to record and mine their transactions in the blockchain, and use a soft handover mechanism to transition the vehicle to another overlay node as it moves in the road network. 
For establishing liability after an accident, we proposed a BlockChain based Framework for Auto-insurance Claim and Adjudication~(B-FICA)~\cite{B-FICA} that uses permissioned blockchain to record all events and interactions of connected and autonomous vehicles. B-FICA includes two partitions, the operational and decision partitions, that record the vehicular events and take decisions on liability respectively. The operational partition includes manufacturers, insurance companies, software providers, service technicians, and the vehicle owners, so that all events are mined into the blockchain by consensus to prevent any one entity from altering the recorded sequence of events. The decision partition includes road and legal authorities, as well as the insurance companies, to consider all available evidence from the operational partition, and make liability decisions. The separation into two partitions ensures that information is disclosed to relevant entities on a need to know basis in each partition. 

The connected vehicles are responsible for producing a significant volume of data. To manage the provided data, SpeedyChain~\cite{SpeedyChain} defines two different sections for each block, the headers and payload. The header contains the basic identification information, while the payload is responsible for keeping the transactions, and the payload is decoupled, which allows the solution to store it externally from the blockchain.

\paragraph{Lessons Learned}
We learned that blockchain architectures for vehicular applications must, in addition to being scalable, explicitly address the challenge of vehicle mobility, whether it is for overlay node handover, maintenance of per-vehicle ledgers, or liability attribution based on spatial and temporal proximity. 

\subsection{Distributed energy trading}
We have also explored how blockchain can enable peer-to-peer energy trading. Renewable energy generation, through solar, wind, and hydro, is experiencing significant growth, which is transforming the energy ecosystem from the simple model of producers and consumers towards having prosumers that can both produce and consume energy. This has led to the emergence of energy trading where prosumers can sell their excess energy to other prosumers or consumers.  

\paragraph{Challenges}
Current energy trading paradigms rely on using the energy company or another trusted third party to act as broker between the seller and buyer of energy \cite{BlockchainTTPenergy}. This centralises interactions, influence, and more importantly trust at the broker, and typically delivers sub-optimal returns to prosumers, due to fees and constraints placed by the broker. It also involves privacy risks for end users, for instance by exposing their energy production and consumption data,  due to the centralisation at the broker. There has been some work on decentralised energy trading with blockchain. In most cases, the energy is still tokenised by a trusted third party \cite{NRGCoin} so that the prosumers can engage in peer-to-peer trading. While this model creates additional separation between the third party and the immediate trading, it still involves centralisation and privacy risks as the token-issuing entity can still make behavioural inferences about the end users based on their token activities. Existing approaches also involves significant blockchain overheads due to the need to broadcast negotiation messages in the network and to mine three transactions into the chain for buyer and seller negotiation. In sum, the challenges of current decentralised energy trading approaches with blockchain were: (1) lack of privacy; (2) reliance on trusted third parties; and (3) blockchain overheads.  

\paragraph{Solutions}
We set out to address these challenges and proposed a Secure and Private Blockchain-based (SPB) energy trading~\cite{SPB} that introduced anonymous backbone routing (ABR) to deal with privacy and communication overhead, atomic metatransactions as part of fully distributed trading process, and a private authentication method to verify smart meters. As for other applications, we used public keys as user identities. Users can adopt multiple public keys to increase their privacy and to avoid linking attacks. Public key-based identities underpin ABR. Instead of using network addresses as node identifiers, ABR routes energy trading messages based on public keys. Backbone ABR nodes are each responsible for routing public keys in a certain prefix range, where the prefix can be set dynamically based on network load. Nodes can then associate each of their PKs with the relevant backbone ABR node, and send their energy trading messages to that ABR node. The messages are then routed through the backbone node that is responsible for the destination node's PK, and then to the destination node itself. Because of the decoupling of the PK from the network address and the node's logical topology location, ABR helps protect node privacy. ABR's unicast rather than broadcast nature avoids bandwidth inefficiencies. 

SPB introduces the concept of atomic meta-transactions to reduce the blockchain overhead and decentralise energy trading. Conventional energy trading based mine three transactions for every trade negotiation into the blockchain \cite{SPB}, which incurs storage overhead. An unsolved problem also involves the buyer and seller dilemma, which requires assurances to both buyer and sellers of an asset or a service that the agreed terms of the trade have been honored. We introduced the concept atomic meta-transactions that are only valid and mined into the blockchain upon the completion of two or more coupled transactions.  Atomic metatransactions in SPB combine the buyer's commit to pay transaction, which places a hold on the funds for energy purchase, and the energy received transaction that confirms successful energy transfer. By mining only one transaction for this trade, SPB reduces storage overhead. To preserve the privacy of the node issuing the energy received transaction, we introduced a certificate of existence, which is created as follows.  Each meter creates a number of keys and forms a Merkle tree of PKs. The meter then sends the root hash of the Merkle tree to another meter to be signed. The signed root is then used as Certificate of Existence (COE), and preserves the original meter's anonymity as it is decoupled from that meter's identity. 

\paragraph{Lessons Learned}
Realising distributed energy trading has intertwined requirements, ranging from meter anonymity to scalability. SPB uses multiple PKs per meter, anonymous backbone routing, atomic meta-transactions, and certificates of existence to jointly meet these requirements. 

\subsection{IoT data markets}
IoT presents an enormous opportunity to transform society by unlocking and unleashing a world of data that, until now, has either been uncollected or has sat largely unused. Application areas where interest in IoT data streams is growing range from health care to personal fitness, smart cities [3], optimization of energy consumption in premises, and many more. For example, in a public transport network, the density of personal travel card swipes over time at individual transit stops may be useful not only to the transportation authority, but also to taxi companies, for better scheduling of their fleet. A new business model referred to as data marketplace~\cite{Perera2013,Misura2016} is thus emerging whereby data producers can sell their IoT data to interested consumers. 

\paragraph{Challenges}

Designing and developing a data marketplace model requires a holistic approach driven by computing and economics to ensure sustainability and usability of the marketplace by increasing number of people adopting it as a platform for buying and selling IoT data. Economic aspects include users' utility maximization and pricing while computing aspects include marketplace key functionalities such as data discovery, negotiation, agreement formation, settlements, data delivery, privacy mechanisms, etc.

A  key aspect that the conventional data marketplaces tend to ignore is ensuring fair trade. The involved parties - sellers, buyers and mediators, being strategic players, may collude and cheat in order to further their gains. Also, if the mediator manages the entire trade, and buyer and seller have no communication, then the mediator is in a position of power. It can falsify information for monetary gains.

Another important issue that needs to be addressed are regulatory and privacy concerns related to data sharing. Different countries have different privacy laws and it is thus necessary to ensure that trade transactions are in compliance. For example, healthcare data needs to comply with Health Insurance Portability and Accountability Act (HIPAA) in USA \cite{HIPAA} and the Privacy Act of 1988 in Australia \cite{PrivacyAct}.

\paragraph{Solutions}

Early works~\cite{Schmid2016}\cite{Cao2016}\cite{Misura2016} exploring this idea rely on the cloud for managing all aspects of the marketplace and thus suffer from known issues such as scalability, expensive infrastructure and single point of failure that plague centralized systems. Recently,  distributed approaches that leverage blockchain technology for designing IoT data marketplaces have been proposed~\cite{Ozyilmaz2018}\cite{Ramachandran2018}. In these IoT marketplace designs, blockchain is used as a distributed and immutable data registry to store the metadata of the IoT data, whereas the IoT data is stored in a distributed database.  

In~\cite{Gupta2018}, we proposed a decentralized IoT data marketplace consisting of IoT data providers, consumers, and multiple trustless brokers. Our marketplace design leverages the smart contract functionality of blockchains to automate the enforcement of the terms of the agreements between data providers and consumers without involving any centralized intermediaries. The smart contract-based data trading scheme uses two types of smart contracts: (1) data subscription contracts maintain the details of subscriptions for provider-consumer pairs, and provide functions for executing, adding, updating and removing subscriptions, and (2) register contract maintains the information about data subscription contracts in a contract lookup table.

\paragraph{Lessons Learned}
There is no trusted intermediary to control the trade, and the market participants may behave maliciously. To minimize the impact of malicious behaviours, data providers send data in batches and consumers pay for receiving each batch rather than paying the full amount after the whole data transfer. Furthermore, to prevent malicious behaviours, trading entities deposit a broker fee amount in an escrow. If an entity behaves maliciously, it is penalized using the amount in the escrow. 

\subsection{Other Applications}
In~\cite{TrustArchitecture}, we proposed a layered architecture that can be used in various CPS applications involving physical observations being stored on blockchains (e.g. healthcare, industrial IoT, social media anaysis, etc.). As a use case scenario, we considered indoor target localization for a smart construction application, where IoT sensors collect data from the construction site to monitor all stages of the construction project. Then, the sensor data is sent to gateway nodes that store the data on a blockchain.  

\paragraph{Challenges}
The IoT sensors collect data from the physical environment so the capture of the IoT data is susceptible to noise, bias, sensor drift, or manipulation by malicious entities. The network should have resistance against inaccurate sensor data. Furthermore, the network may consist of a large number of IoT devices frequently observing the environment and generating a large amount of data. The architecture should be able to handle the amount of data generated at the data layer.

At the blockchain layer, the gateway nodes that receive data from IoT sensors and record it on the blockchain may be malicious. Through the block verification mechanism, invalid blocks created by the malicious gateway nodes should be detected and discarded. Besides, the overhead caused by the block verification mechanism should be low.

\paragraph{Solutions}
To address these challenges, we proposed a data trust and gateway reputation module, lightweight block generation, adaptive block validation, and distributed consensus mechanisms in~\cite{TrustArchitecture}. Our data trust module evaluates the trustworthiness of a sensor observation based on the confidence of the sensor, evidence from other sensor observations, and reputation of the sensor node. 

Since the identities of the gateway nodes are known and they have permissions to participate in the blockchain network, they do not need to compete with each other for block mining using computationally expensive mechanisms. Instead, the gateway nodes use a lightweight block generation mechanism to create a new block by grouping sensor observations together in periodic intervals. Our gateway reputation module calculates the reputations of gateway nodes. These reputation values are used by the adaptive block validation mechanism to determine the percentage of transactions to be validated in a block by each validator node. The higher the number of validators and the higher the reputation of the block generating node, the lower the percentage of transactions to be chosen randomly and validated by each validator and the lower the overhead of block validation.   

\paragraph{Lessons Learned}
We learned that for improving the end-to-end trust in CPS applications involving physical observations being stored on blockchains, we need to take into account the trustworthiness of sensor data and the blockchain nodes. Evaluating the trustworthiness of the blockchain nodes let us adapt the block validation mechanism and improve the system performance.  

\section{Discussion}

In this paper, we presented our research outcomes and experiences in applying blockchain technology for CPS. Our journey started by recognizing blockchain as a promising technology that can address the CPS challenges, such as security, privacy, centralization, resource constraints, scalability, lack of control and auditability, and complex interactions. However, adopting blockchain for CPS is not straightforward either and has its own challenges. Low scalability, high latency, low throughput, computationally expensive consensus mechanisms, trust and privacy related problems of blockchains are significant barriers to blockchain adoption for CPS. Thus, we started our research in blockchain for CPS by developing mechanisms to address these common problems and building blockchains that are appropriate for CPS. Building on the mechanisms that we developed, we later started our research on CPS applications that have unique constraints and requirements, such as smart cities and smart homes, autonomous vehicles and liability, supply chains, energy trading, IoT data markets, and smart construction. As blockchain is an emerging technology with a potential to improve the performance of CPS, many research questions and opportunities exist in the areas of designing novel blockchain mechanisms for CPS and adopting blockchains for CPS applications.    

A current problem for the design of blockchain-based solutions for CPS is the use of oversimplified system models that do not capture the system dynamics, requirements, and constraints well. More realistic system models, real world data sets, proof-of-concept and pilot trials are required for designing blockchain-based solutions that can be applied to real-life CPS. Another important challenge is the lack of evaluation tools to compare the performance of alternative solutions. Although some blockchain platforms offer testnets, and performance evaluation tools, they are not designed to evaluate the performance of large scale networks and complex CPS applications. Thus, there is a need to design tools that can evaluate the performance of large scale networks and complex CPS applications effectively.

\section*{Acknowledgments}
The work has been supported by the Cyber Security Research Centre Limited whose activities are partially funded by the Australian Government’s Cooperative Research Centres Programme.


\begin{thebibliography}{11}

\bibitem{Rajkumar2010} Ragunathan Rajkumar,  Insup Lee, Lui Sha, John Stankovic, "Cyber-physical systems: The next computing revolution," Design Automation Conference, Anaheim, CA, 2010, pp. 731-736.

\bibitem{BlockchainInIoT} Ali Dorri, Salil S. Kanhere, and Raja Jurdak, "Blockchain in internet of things: challenges and solutions," arXiv preprint arXiv:1608.05187 (2016).

\bibitem{Satoshi2008} N. Satoshi, "Bitcoin: A peer-to-peer electronic cash system," 2008.

\bibitem{DependableIoT} Avelino F. Zorzo, Henry C. Nunes, Roben C. Lunardi, Regio A. Michelin, and Salil S. Kanhere, "Dependable IoT using blockchain-based technology," In 2018 Eighth Latin-American Symposium on Dependable Computing (LADC), pp. 1-9. IEEE, 2018.

\bibitem{Khaitan2015} Siddhartha K. Khaitan, James D. McCalley, "Design Techniques and Applications of Cyberphysical Systems: A Survey," in IEEE Systems Journal, vol. 9, no. 2, pp. 350-365, June 2015.

\bibitem{LSB} Ali Dorri, Salil S. Kanhere, Raja Jurdak, Praveen Gauravaram,
"LSB: A Lightweight Scalable Blockchain for IoT security and anonymity," Journal of Parallel and Distributed Computing, Volume 134, 2019, Pages 180-197

\bibitem{SpeedyChain} Regio A. Michelin, Ali Dorri, Marco Steger, Roben C. Lunardi, Salil S. Kanhere, Raja Jurdak, and Avelino F. Zorzo, "SpeedyChain: A framework for decoupling data from blockchain for smart cities," In Proceedings of the 15th EAI International Conference on Mobile and Ubiquitous Systems: Computing, Networking and Services, pp. 145-154. ACM, 2018.

\bibitem{SPB} Ali Dorri, Fengji Luo, Salil S. Kanhere, Raja Jurdak, and Zhao Yang Dong, "SPB: A Secure Private Blockchain-Based Solution for Distributed Energy Trading," IEEE Communications Magazine 57, no. 7 (2019): 120-126.

\bibitem{B-FICA} Chuka Oham, Raja Jurdak, Salil S. Kanhere, Ali Dorri, and Sanjay Jha, "B-fica: Blockchain based framework for auto-insurance claim and adjudication," In 2018 IEEE iThings and IEEE GreenCom and IEEE CPSCom and IEEE SmartData, pp. 1171-1180. IEEE, 2018.

\bibitem{ProductChain} Sidra Malik, Salil S. Kanhere, and Raja Jurdak, "Productchain: Scalable blockchain framework to support provenance in supply chains." In 2018 IEEE 17th International Symposium on Network Computing and Applications (NCA), pp. 1-10. IEEE, 2018.

\bibitem{TrustChain} Sidra Malik, Volkan Dedeoglu, Salil S Kanhere, and Raja Jurdak, "TrustChain: Trust Management in Blockchain and IoT supported Supply Chains." IEEE International Conference on Blockchain (Blockchain 2019), 2019.

\bibitem{TrustArchitecture} Volkan Dedeoglu, Raja Jurdak, Guntur D. Putra, Ali Dorri, and Salil S. Kanhere, "A Trust Architecture for Blockchain in IoT," In 16th EAI International Conference on Mobile and Ubiquitous Systems (Mobiqutious 2019), Houston, USA, November 2019.
\bibitem{Ozyilmaz2018} Kazim Rifat Ozyilmaz, Mehmet Dogan , Arda Yurdakul, "IDMoB IoT Data Marketplace on Blockchain," 2018 Crypto Valley Conference on Blockchain Technology (CVCBT), Zug, 2018, pp. 11-19.

\bibitem{Ramachandran2018} Gowri S. Ramachandran, Rahul Radhakrishnan, Bhaskar Krishnamachari, "Towards a Decentralized Data Marketplace for Smart Cities", 2018 IEEE International Smart Cities Conference (ISC2), Kansas City, MO, USA, 2018, pp. 1-8.

\bibitem{Perera2013} Charith Perera, Arkady Zaslavsky, Peter Christen, and Dimitrios Georgakopoulos,“Sensing as a service model for smart cities supported by Internet of Things,” Transactions on Emerging Telecommunications Technologies, vol. 25, no. 1, pp. 81–93, Feb. 2013.

\bibitem{Misura2016} Kresimir Misura and Mario Zagar, “Data marketplace for Internet of Things,” 2016 International Conference on Smart Systems and Technologies (SST), 2016.

\bibitem{Gupta2018} Pooja Gupta, Salil S. Kanhere, Raja Jurdak, "A Decentralized IoT Data Marketplace," In proceedings of the 3rd Symposium on Distributed Ledger Technology, Gold Coast, Australia, November 2018.
\bibitem{MOF-BC} Ali Dorri, Salil S. Kanhere, and Raja Jurdak, "MOF-BC: A memory optimized and flexible blockchain for large scale networks," Future Generation Computer Systems 92, 2019, pp. 357-373.
\bibitem{Angelis} Jannis Angelis, and Elias Ribeiro da Silva, "Blockchain adoption: A value driver perspective," Business Horizons, vol. 62, no. 3, pp. 307-314, 2019.
\bibitem{Privacy-Preserving} Ali Dorri, Clemence Roulin, Raja Jurdak, and Salil S. Kanhere, "On the Activity Privacy of Blockchain for IoT," 2018, arXiv preprint arXiv:1812.08970.

\bibitem{Wang2018} Shuai Wang, Yong Yuan, Xiao Wang, Juanjuan Li, Rui Qin, Fei-Yue Wang, "An Overview of Smart Contract: Architecture, Applications, and Future Trends," 2018 IEEE Intelligent Vehicles Symposium (IV), Changshu, 2018, pp. 108-113.

\bibitem{Krzanich} Brian Krzanich, "Data is the New Oil in the Future of Automated Driving," Available at https://newsroom.intel.com/editorials/krzanich-the-future-of-automated-driving/, Accessed: 04 nov 2019.

\bibitem{Schmid2016} Stefan Schmid, Arne Bröring, Denis Kramer, Sebastian Käbisch, Achille Zappa, Martin Lorenz, Yong Wang, Andreas Rausch, and Luca Gioppo. "An architecture for interoperable IoT Ecosystems." In International Workshop on Interoperability and Open-Source Solutions, pp. 39-55. Springer, Cham, 2016.

\bibitem{Cao2016} Tien-Dung Cao, Tran-Vu Pham, Quang-Hieu Vu, Hong-Linh Truong, Duc-Hung Le, and Schahram Dustdar, "MARSA: A marketplace for realtime human sensing data," ACM Transactions on Internet Technology (TOIT) 16, no. 3 (2016): 16.

\bibitem{Illegal2019} Samuel Gibbs, "Child abuse imagery found within bitcoin's blockchain," The Guardian, 20 March 2018, Retrieved from https://www.theguardian.com
 
\bibitem{BCAutomotive2019} Ali Dorri, Marco Steger, Salil S. Kanhere, and Raja Jurdak, "Blockchain: A distributed solution to automotive security and privacy," IEEE Communications Magazine 55, no. 12 (2017): 119-125.

\bibitem{BlockchainTTPenergy} Zhetao Li, Jiawen Kang, Rong Yu, Dongdong Ye, Qingyong Deng, and Yan Zhang, "Consortium blockchain for secure energy trading in industrial internet of things," IEEE transactions on industrial informatics 14, no. 8 (2017): 3690-3700.

\bibitem{NRGCoin} Mihail Mihaylov, Ivan Razo-Zapata, and Ann Nowe, "NRGcoin—A Blockchain-based Reward Mechanism for Both Production and Consumption of Renewable Energy," In Transforming Climate Finance and Green Investment with Blockchains, pp. 111-131. Academic Press, 2018.

\bibitem{GDPR_reg} European Parliament, Council of the European Union, "Regulation (EU) 2016/679 of the European Parliament and of the Council of 27 April 2016 on the protection of natural persons with regard to the processing of personal data and on the free movement of such data, and repealing Directive 95/46/EC (General Data Protection Regulation) (Text with EEA relevance)," OJ L 119, 4.5.2016, p. 1–88

\bibitem{HIPAA} Act, Accountability. "Health insurance portability and accountability act of 1996." Public law 104 (1996): 191.

\bibitem{PrivacyAct} The Privacy Act 1988 (Cwlth),

Retrieved from https://www.legislation.gov.au/Series/C2004A03712

\bibitem{SmartHomeDataset} https://iotanalytics.unsw.edu.au/
\end{thebibliography}
\end{document}